\documentclass[10pt]{article}
\usepackage[margin=1in]{geometry}
\usepackage{latexsym,amssymb,amsmath,amsfonts,xcolor,cases,mathtools}
\usepackage{caption,graphicx,theorem}
\usepackage{subfig,wrapfig}
\usepackage[pagewise]{lineno}
\usepackage{epstopdf,chngcntr}
\usepackage{natbib}
\usepackage{hyperref}
\usepackage{stfloats}

\numberwithin{equation}{section}

\hypersetup{
    colorlinks = true,
    linkcolor = {blue},
    citecolor = {blue},
    urlcolor  = {blue},}

\newcommand{\IM}[1]{\textup{Im}\left(#1\right)}
\newcommand{\CC}{{\mathbb{C}}}

\newcommand{\ZZ}{{\mathbb{Z}}}

\newcommand{\Der}[2]{\frac{\text{d}#1}{\text{d}#2}}

\newcommand{\mathd}{\mathrm{d}}

\newcommand{\BF}[1]{\mathbf{#1}}

\definecolor{UoM}{RGB}{100,30,160}
\definecolor{newgreen}{RGB}{0,150,0}

\begin{document}
\title{Scattering by two parallel semi--infinite arrays of sound--soft scatterers}

\author{M. A. Nethercote$^*$\\
$^*$ mn598@cam.ac.uk}

\date{}
\maketitle

\begin{abstract}
This article considers the scattering of an acoustic wave by two identical semi--infinite arrays of point scatterers which are positioned parallel to one another. The resulting system of scattering coefficients is formulated as a pair of coupled discrete Wiener--Hopf equations with a full $2\times2$ matrix kernel composed of two scalar kernel functions that represent interaction within the selected array and between the two arrays. Exploiting the symmetry of the two arrays, we introduce symmetric and antisymmetric components of the scattering coefficients, which decouple the matrix Wiener--Hopf system into two independent scalar problems with kernels given by the sum and difference of the two original kernel functions. The scalar Wiener--Hopf equations are then solved using the Wiener--Hopf technique in a very similar way to what has previously been used to solve the semi--infinite array problem. The original definitions of the two scalar kernels are represented by a lattice sum and practical evaluation techniques are discussed. Comparisons with previous work are also critically discussed. The resulting formulation provides an exact analytical solution to a coupled semi--infinite array problem and illustrates how symmetry can reduce a discrete matrix Wiener--Hopf problem to two scalar problems.
\end{abstract}

\section{Introduction}
The application of analytical methods for scattering problems involving infinite and semi-infinite structures remains a relevant field in modern research, with the Wiener--Hopf technique \citep{WienerHopf1931} being a powerful and popular method with a rich history \citep{LawrieAbrahams2007}. In the context of continuous parallel semi-infinite structures, \citet{Heins1948part1,Heins1948part2} obtained exact Wiener--Hopf solutions for the reflection and transmission coefficients of a pair of semi-infinite parallel plates. These works provide an early example in which the symmetry of two parallel structures can be exploited to simplify a coupled scattering problem. Subsequent work by \citet{Abrahams1987a} considered the acoustic scattering of sound by two parallel semi-infinite screens pointing in opposite directions, while \citet{Abrahams1988,Abrahams1988part2} developed matrix Wiener--Hopf formulations and explicit factorisations for acoustic scattering by two semi-infinite plates which were parallel but staggered. These studies established a substantial body of work on coupled semi-infinite structures and illustrate the additional analytical difficulties introduced by matrix Wiener--Hopf kernels. More generally, constructive and approximate approaches to matrix Wiener--Hopf factorisation are discussed by \citet{KisilWHReview2021} and \citet{RogosinMishuris2016}.

A separate but closely related line of research concerns scattering by periodic arrays of point scatterers. In particular, \citet{HillsKarp1965} and \citet{LintonMartin2004} developed a unified Wiener--Hopf treatment of a semi-infinite array of isotropic point scatterers, establishing a discrete Wiener--Hopf formulation for the corresponding infinite system of scattering coefficients. Following this, \citet{LintonPorterThompson2007} also studied scattering by semi-infinite arrays with particular consideration on the possible presence of Rayleigh-Bloch waves and the special case of resonance. These works provide the basis for applying discrete Wiener--Hopf methods to semi-infinite arrays. The associated lattice sums are themselves an important component of periodic scattering theory and \citet{Linton1998,Linton2006,Linton2010} considered the Green's function and lattice sums arising in many types of periodic problems. Meanwhile, \citet{Lynott_19} developed efficient methods of tail-end asymptotics for the quasi-periodic Green's function associated with an infinite array, which has proven to be useful in the practical evaluation of slow-convergent series involving Hankel functions. The use of Foldy's approximation \citep{Foldy1945} in periodic scattering problems has been considered in many similar settings with small, sound-soft scatterers, including infinite lattices \citep{KrynkinMcIver2009} and resonance in single arrays \citep{ArrayResonanceFoldyPaper}.

Coupling between multiple arrays introduces additional structure and complexity to the resulting scattering problem. \citet{Chremmos2011} investigated travelling electromagnetic waves in two parallel infinite arrays of point scatterers (dubbed nanoparticle chains), demonstrating the importance of coupling between arrays and the associated symmetric and antisymmetric wave components. For semi-infinite arrays, \citet{HWpaperI} considered acoustic diffraction by a wedge of point scatterers and developed a method using the discrete Wiener--Hopf technique for the resulting coupled array problem. This approach was later generalised by \citet{MSIApaper} to an arbitrary number of periodic semi-infinite arrays with arbitrary positioning, orientation and properties. In that formulation, the scattering coefficients satisfy coupled systems of equations which are solved using the discrete Wiener-Hopf technique which leads to a matrix equation for the coefficients. The general formulation provides a framework for multiple-array configurations, but the resulting matrix problem is considerably more complicated than the scalar Wiener-Hopf problem associated with a single array. 

In elastic plates, scattering and localisation of flexural waves have also been studied using discrete Wiener--Hopf methods. \citet{Movchan2009}, \citet{Haslinger2014} and \citet{Haslinger2016} investigated trapped and blocked waves in platonic gratings, while \citet{JonesMovchan2017} considered two parallel semi-infinite gratings of rigid pins on a Kirchhoff plate. \citet{HaslingerJonesMovchan2018} extended this to configurations of multiple parallel semi-infinite gratings, including a herringbone waveguide formed by four gratings. These studies used discrete Wiener--Hopf formulations to identify trapped modes through analysis of the kernel, with the latter employing a dipole approximation for closely spaced pairs of gratings. 

The present work combines these approaches by considering two identical semi-infinite arrays of point scatterers positioned parallel to one another. Pursuing a Wiener--Hopf formulation of the problem leads to a $2\times2$ matrix Wiener--Hopf system with a matrix kernel that is constructed from two unique entries; $K(z)$ as the diagonal entries and $L(z)$ as the off-diagonal entries. Although a general matrix factorisation would be required for arbitrary multiple-array configurations, the symmetry of the present geometry permits an exact decomposition into even and odd components. The resulting problem consists of two independent scalar Wiener--Hopf equations with kernels $K(z)+L(z)$ and $K(z)-L(z)$, respectively. We exploit this reduction to obtain explicit factorisations and scattering coefficients.

This article is structured as follows. We set up the problem and find the system of equations we must solve in Section \ref{sec:setup}. We solve this problem using the discrete analogue of the Wiener--Hopf technique in Section \ref{sec:solve} before finding the scattering coefficients in Section \ref{sec:invZT}. Some numerical results and test cases are displayed in Section \ref{sec:results}. Appendix \ref{sec:Kernels} gives some more details on the practical evaluation of the kernel functions encountered in this article.

\section{Setting up the problem}\label{sec:setup}
\begin{figure}[ht]\centering
\includegraphics[width=0.6\textwidth]{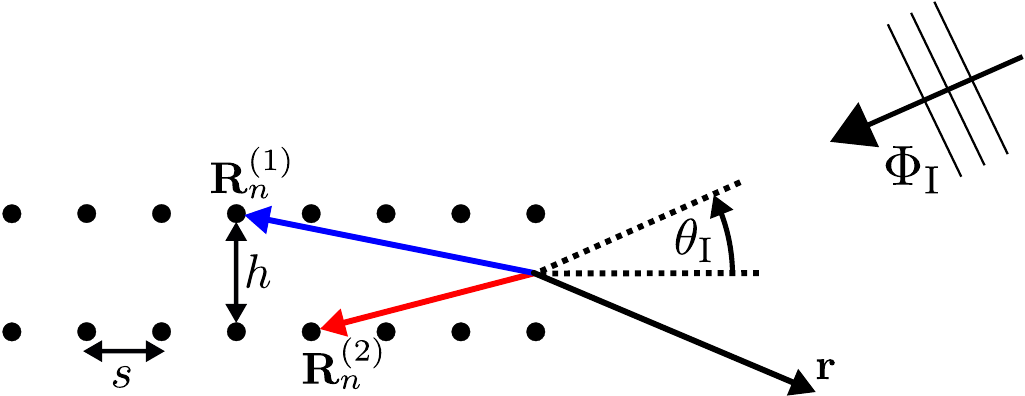}
\caption{Diagram of the two parallel arrays, separated by height $h$, which are constructed from identical sound-soft cylinders with infinite height, radius $a$ and spacing $s$. We also include the incident wave $\Phi_{\textrm{I}}$ with incident angle $\theta_{\textrm{I}}$ as well as the position vectors $\BF{r}$, $\BF{R}^{(1)}_n$ and $\BF{R}^{(2)}_n$.}
\label{fig:ParaA-diagram}
\end{figure}

In three dimensions, the considered scatterers are all cylinders of infinite height, have a small radius and are sound-soft, i.e. satisfy homogeneous Dirichlet boundary conditions. The direction of the incoming incident disturbance will be perpendicular to the cylinders, which implies normal (non-skew) incidence. This means that the problem is naturally reduced to two dimensions. This article is considering a specific problem with two identical and parallel semi-infinite arrays, separated by a height $h>0$ and constructed from scatterers which have radius $a>0$ and are spaced apart by $s>0$. This means that the scatterer locations are given by 
\begin{align}\label{ParaA-Rn}
\BF{R}^{(1)}_n=-ns\BF{\hat{x}}+\frac{h}{2}\BF{\hat{y}},\quad\BF{R}^{(2)}_n=-ns\BF{\hat{x}}-\frac{h}{2}\BF{\hat{y}}
\end{align}
for the top and bottom array respectively where $(\BF{\hat{x}},\BF{\hat{y}})$ are the basis unit vectors of a Cartesian coordinate system, the origin of which is placed exactly halfway between the first scatterers of each array. We will need $2a<\text{min}(s,h)$ to prevent the scatterers from intersecting and overlapping. We are looking for time-harmonic solutions to the linear wave equation by assuming and then suppressing the time factor $e^{-i\omega t}$, where $\omega$ is the angular frequency. We use a polar coordinate system $(r,\theta)$ with the position vector $\BF{r}$ and let $\Phi$ be the total wave field which includes the addition of a unit amplitude plane wave incident at angle $\theta_{\textrm{I}}$ given by,
\begin{align}\label{ParaA-PhiI}
\Phi_{\textrm{I}}=e^{-ikr\cos(\theta-\theta_{\textrm{I}})},
\end{align}
and the scattered wave field $\Phi_{\textrm{S}}$. Each of these fields satisfy the Helmholtz equation with wavenumber $k$. Figure \ref{fig:ParaA-diagram} displays a diagram of this problem including the two parallel arrays, the incident wave and the position vectors of the scatterers.

We use Foldy's approximation which assumes that the ratio of the scatterer size with the wavelength is small, i.e. $ka\ll1$. Therefore the scatterers are assumed to be isotropic, which implies that the scattered field can be written in the form of a monopole expansion so the total field is given by,
\begin{align}\label{ParaA-A-gensol}
\Phi=\Phi_{\textrm{I}}+\sum_{n=0}^\infty\left[A^{(1)}_nH^{(1)}_0(k|\BF{r}-\BF{R}^{(1)}_n|)\right]+\sum_{n=0}^\infty\left[A^{(2)}_nH^{(1)}_0(k|\BF{r}-\BF{R}^{(2)}_n|)\right].
\end{align}
Here, $A^{(1)}_n$ (resp. $A^{(2)}_n$) are the scattering coefficients associated with the $n^{\textrm{th}}$ scatterer of the top (resp. bottom) array. We use the same procedure as described in \citep{MSIApaper} which is to apply the sound-soft boundary conditions and approximate for small scatterers. As a result, we obtain two systems of equations for $A^{(1)}_m$ and $A^{(2)}_m$ for $m\geq0$,
\begin{align}\nonumber\label{ParaA-A-system}
A^{(1)}_mH^{(1)}_0(ka)+&\sum_{\substack{n=0\\n\neq m}}^{\infty}\left[A^{(1)}_nH^{(1)}_0(ks|m-n|)\right] +\sum_{n=0}^{\infty}\left[A^{(2)}_nH^{(1)}_0\left(k\Lambda(m-n)\right)\right]=-e^{iksm\cos(\theta_{\text{I}})-\frac{ikh}{2}\sin(\theta_{\text{I}})},\\
A^{(2)}_mH^{(1)}_0(ka)+&\sum_{\substack{n=0\\n\neq m}}^{\infty}\left[A^{(2)}_nH^{(1)}_0(ks|m-n|)\right] +\sum_{n=0}^{\infty}\left[A^{(1)}_nH^{(1)}_0\left(k\Lambda(n-m)\right)\right]=-e^{iksm\cos(\theta_{\textrm{I}})+\frac{ikh}{2}\sin(\theta_{\text{I}})},
\end{align}
where $\Lambda(\ell)=\sqrt{(s\ell)^2+h^2}$. We have remarked in \citep{MSIApaper} that using $H^{(1)}_0(ka)$ can satisfy energy conditions to leading order. \citep{LintonMartin2004} used the leading order approximation of $H^{(1)}_0(ka)$ instead but this does not change the solution to leading order. 

To solve this system of equations \eqref{ParaA-A-system}, we use the discrete analogue of the Wiener-Hopf technique in a very similar manner to \citep{LintonMartin2004}. Noting that $A^{(j)}_m=0$ if $m<0$, we start by extending the summations to $n<0$ and extend the definition of the right hand side to be equal to some unknown coefficients $F^{(j)}_m$ when $m<0$.
\begin{align}\nonumber\label{ParaA-A-system-extend}
A^{(1)}_mH^{(1)}_0(ka)+&\sum_{\substack{n=-\infty\\n\neq m}}^{\infty}\left[A^{(1)}_nH^{(1)}_0(ks|m-n|)\right] +\sum_{n=-\infty}^{\infty}\left[A^{(2)}_nH^{(1)}_0\left(k\Lambda(m-n)\right)\right]\\
=&\begin{cases}-e^{iksm\cos(\theta_{\text{I}})-\frac{ikh}{2}\sin(\theta_{\text{I}})}& m\geq0,\\ F^{(1)}_m& m<0,\end{cases}\\
\nonumber A^{(2)}_mH^{(1)}_0(ka)+&\sum_{\substack{n=-\infty\\n\neq m}}^{\infty}\left[A^{(2)}_nH^{(1)}_0(ks|m-n|)\right] +\sum_{n=-\infty}^{\infty}\left[A^{(1)}_nH^{(1)}_0\left(k\Lambda(n-m)\right)\right]\\
=&\begin{cases}-e^{iksm\cos(\theta_{\textrm{I}})+\frac{ikh}{2}\sin(\theta_{\text{I}})}& m\geq0,\\ F^{(2)}_m& m<0,\end{cases}
\end{align}

\begin{figure}[ht]\centering
\includegraphics[width=0.5\textwidth]{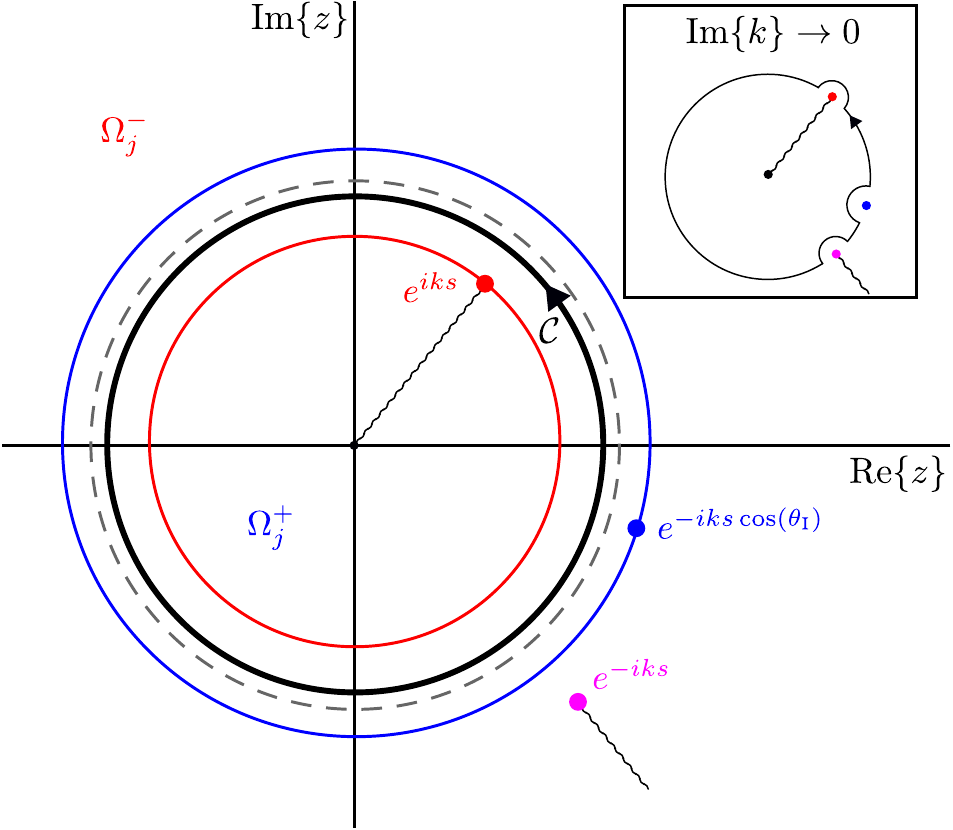}
\caption{Diagram of the regions $\Omega^\pm$ on the $z$ complex plane. The border of the regions $\Omega^\pm$ are shown as blue and red circles respectively, the gray dashed circle is the unit circle $|z|=1$ and the solid black line is the integration contour $\mathcal{C}$. The smaller diagram is the special case where $\IM{k}\rightarrow0$.}
\label{fig:circle_contour}
\end{figure}

Consider the Z--transform and the associated inverse given by,
\begin{align}\label{ParaA-Z-transform}
G(z)=\sum_{m=-\infty}^{\infty} G_mz^m,\quad
G_m=\frac{1}{2\pi i}\int_{\mathcal{C}}G(z)z^{-m-1}\mathd z.
\end{align}
where the inverse contour $\mathcal{C}$ runs anticlockwise about the unit circle. After we apply the Z--transform to the systems \eqref{ParaA-A-system-extend}, we obtain two coupled Wiener-Hopf equations,
\begin{align}
\label{ParaA-A-WHE1}K(z)A^+_1(z)+L(z)A^+_2(z)&=F^+_{\text{pole},1}(z)+F^-_1(z),\\
\label{ParaA-A-WHE2}K(z)A^+_2(z)+L(1/z)A^+_1(z)&=F^+_{\text{pole},2}(z)+F^-_2(z),
\end{align}
where the unknowns to find are $A^+_j(z)$ defined by
\begin{align}\label{ParaA-A+def}
A_j(z)=\sum_{m=-\infty}^{\infty} A^{(j)}_mz^m=\sum_{m=0}^{\infty} A^{(j)}_mz^m=A^+_j(z).
\end{align}
These Wiener-Hopf equations have two kernel functions $K(z)$ and $L(z)$,
\begin{align}
\label{ParaA-K(z)}K(z)&=H^{(1)}_0(ka)+\sum_{\ell=1}^\infty \left[(z^\ell+z^{-\ell})H^{(1)}_0(ks\ell)\right],\\
\label{ParaA-L(z)}L(z)&=\sum_{\ell=-\infty}^\infty \left[z^\ell H^{(1)}_0\left(k\sqrt{(s\ell)^2+h^2}\right)\right]
\end{align}
and the forcing terms are given by,
\begin{align}
\label{ParaA-A-Fpole}F^+_{\text{pole},1}(z)&
=\frac{e^{-\frac{ikh}{2}\sin(\theta_{\text{I}})}}{ze^{iks\cos(\theta_{\text{I}})}-1},\quad
F^+_{\text{pole},2}(z)
=\frac{e^{\frac{ikh}{2}\sin(\theta_{\text{I}})}}{ze^{iks\cos(\theta_{\text{I}})}-1},\\
\label{ParaA-A-F-}F^-_1(z)&=\sum_{m=-\infty}^{-1} F^{(1)}_mz^m,\quad F^-_2(z)=\sum_{m=-\infty}^{-1} F^{(2)}_mz^m.
\end{align}
As in \citep{HWpaperI,MSIApaper}, it is useful to assume that $k$ has a small positive part to improve the convergence of the Z--transform. We then define the two regions,
\begin{align}\nonumber\label{ParaA-regions}
\Omega^+&=\left\{z\in\CC: |z|<e^{\IM{k}s\cos(\theta_{\textrm{I}})}\right\},\\
\Omega^-&=\left\{z\in\CC: |z|>e^{-\IM{k}s}\right\},
\end{align}
for which functions with $+$ and $-$ superscripts are analytic in $\Omega^+$ and $\Omega^-$ respectively (see Figure \ref{fig:circle_contour}). 

The Wiener--Hopf kernels $K(z)$ and $L(z)$ have a few similar properties including having branch points at $e^{\pm iks}$ with associated branch cuts going to zero and complex infinity respectively and additionally will both satisfy the identity $K(1/z)=K(z)$. Both are also analytic and zero-free on the annulus,
$$\left\{z\in\CC: e^{-\IM{k}s}<|z|<e^{\IM{k}s}\right\},$$
which contains the intersection region $\Omega^+\cap\Omega^-$. See Appendix \ref{sec:Kernels} for some practical notes on evaluating these kernel functions. This can be done by using the method of tail-end asymptotics or by finding alternative formulae. To solve the Wiener-Hopf equations \eqref{ParaA-A-WHE1} and \eqref{ParaA-A-WHE2} directly, we must know how to factorise the matrix Wiener--Hopf kernel
\begin{align*}
\begin{pmatrix}
K(z)&L(z)\\L(1/z)&K(z)
\end{pmatrix},
\end{align*}
however this is going to be very difficult to factorise without using approximate methods \citep{KisilWHReview2021}. Alternatively, we can take inspiration from \citep{Heins1948part1,Heins1948part2} and consider the even and odd components of the coefficients by the definitions, 
\begin{align}\label{ParaA-BC-def}
B_m=\frac{A^{(1)}_m+A^{(2)}_m}{2}\quad \textrm{and}\quad C_m=\frac{A^{(1)}_m-A^{(2)}_m}{2}.
\end{align}
We can obtain a new matrix Wiener-Hopf problem using similar steps to \eqref{ParaA-BC-def},
\begin{align}\label{ParaA-WHE-BC}
\begin{pmatrix}
K_B(z)&0\\
0&K_C(z)
\end{pmatrix}
\begin{pmatrix}B^+(z)\\C^+(z)\end{pmatrix}&=
\begin{pmatrix}F^+_{\text{pole},B}(z)\\F^+_{\text{pole},C}(z)\end{pmatrix}
+\begin{pmatrix}F^-_B(z)\\F^-_C(z)\end{pmatrix}
\end{align}
where $B^+(z)$ and $C^+(z)$ are the Z--transforms (with very similar definitions as \eqref{ParaA-A+def}) of the unknown coefficients $B_m$ and $C_m$ respectively. Here, the kernel functions are given by $K_B(z)=K(z)+L(z)$ and $K_C(z)=K(z)-L(z)$, and the forcing terms are given by
\begin{align}
\label{ParaA-B-Fpole}F^+_{\text{pole},B}(z)&=\frac{F^+_{\text{pole},1}(z)+F^+_{\text{pole},2}(z)}{2}
=\frac{\cos\left(\frac{kh}{2}\sin(\theta_{\text{I}})\right)}{ze^{iks\cos(\theta_{\text{I}})}-1},\\
\label{ParaA-C-Fpole}F^+_{\text{pole},C}(z)&=\frac{F^+_{\text{pole},1}(z)-F^+_{\text{pole},2}(z)}{2}
=-\frac{i\sin\left(\frac{kh}{2}\sin(\theta_{\text{I}})\right)}{ze^{iks\cos(\theta_{\text{I}})}-1},\\
\label{ParaA-B-F-}F^-_B(z)&=\frac{F^-_1(z)+F^-_2(z)}{2},\\
\label{ParaA-C-F-}F^-_C(z)&=\frac{F^-_1(z)-F^-_2(z)}{2}.
\end{align}

\subsection{Solving the Wiener--Hopf problem}\label{sec:solve}
The Wiener-Hopf equations in this system are now completely decoupled and can be solved using the same procedure as featured in \citep{LintonMartin2004} for the semi-infinite array problem. Take the first equation of \eqref{ParaA-WHE-BC}. The first step of the Wiener--Hopf technique is to factorise the Wiener-Hopf kernel $K_B(z)$. We would write this factorisation as $K_B(z)=K_B^+(z)K_B^-(z)$ where the two factors are defined by Cauchy's integral formulae \eqref{ParaA-KB^+(z)}. 
\begin{align}\label{ParaA-KB^+(z)}
\nonumber\ln(K_B^+(z))=\ln(K_B^{0})-&\frac{1}{2\pi i}\int_{\mathcal{C}}\frac{\ln(K_B(\xi))}{\xi-1/z}\text{d}\xi,\\
\nonumber\ln(K_B^-(z))=\ln(K_B^{0})-&\frac{1}{2\pi i}\int_{\mathcal{C}}\frac{\ln(K_B(\xi))}{\xi-z}\text{d}\xi,\\
\ln(K_B^{0})=&\frac{1}{4\pi i}\int_{\mathcal{C}}\frac{\ln(K_B(\xi))}{\xi}\text{d}\xi.
\end{align}
Here, the integration contour $C$ is the anticlockwise circular path contained inside $\Omega^+\cap\Omega^-$ on the $\xi$ complex plane (see Figure \ref{fig:circle_contour}). This contour also does not include the poles at $\xi=z^{\pm1}$. Both kernel factors $K_B^\pm(z)$ are also analytic and zero-free inside the regions $\Omega^\pm$ respectively. We define these factors in this way such that they also satisfy $K_B^-(z)=K_B^+(1/z)$. 

Dividing the first equation of \eqref{ParaA-WHE-BC} by $K_B^-(z)$ will give us, 
\begin{align}\label{ParaA-WHE-B-fac}
K_B^+(z)B^+(z)&=\frac{F^+_{\text{pole},B}(z)}{K_B^-(z)}+\frac{F^-_B(z)}{K_B^-(z)},
\end{align}
for which the next step is sum-splitting the pole term $\frac{F^+_{\text{pole},B}(z)}{K_B^-(z)}$ by removing the pole at $z=e^{-iks\cos(\theta_{\textrm{I}})}$,
\begin{align}\label{ParaA-WHE-B-sumsplit}
\frac{F^+_{\text{pole},B}(z)}{K_B^-(z)}=\frac{F^+_{\text{pole},B}(z)}{K_B^-(e^{-iks\cos(\theta_{\textrm{I}})})}
+F^+_{\text{pole},B}(z)\left[\frac{1}{K_B^-(z)}-\frac{1}{K_B^-(e^{-iks\cos(\theta_{\textrm{I}})})}\right],
\end{align}
where the first and second terms of the sum-split are analytic in $\Omega^+$ and $\Omega^-$ respectively. This brings us to the final Wiener-Hopf equation
\begin{align}\label{ParaA-WHE-B-final}
K_B^+(z)B^+(z)-\frac{F^+_{\text{pole},B}(z)}{K_B^-(e^{-iks\cos(\theta_{\textrm{I}})})}&=
F^+_{\text{pole},B}(z)\left[\frac{1}{K_B^-(z)}-\frac{1}{K_B^-(e^{-iks\cos(\theta_{\textrm{I}})})}\right]+\frac{F^-_B(z)}{K_B^-(z)}.
\end{align}
Every term on the left-hand (resp. right-hand) side is analytic in $\Omega^+$ (resp. $\Omega^-$). The two sides of \eqref{ParaA-WHE-B-final} are used to construct a function $\Psi(z)$ defined by
\begin{align}
\Psi(z)=\begin{cases}
\eqref{ParaA-WHE-B-final}(\textrm{LHS}) & z\in\Omega^+,\\
\eqref{ParaA-WHE-B-final}(\textrm{RHS}) & z\in\Omega^-,\\
(\textrm{LHS})=(\textrm{RHS}) & z\in\Omega^+\cap\Omega^-.\\
\end{cases}
\end{align}
This function $\Psi$ is entire and since each term of the right-hand side of \eqref{ParaA-WHE-B-final} is $O\left(\frac{1}{z}\right)$ as $|z|\rightarrow\infty$, we know that $\Psi$ is bounded and decays to zero at infinity. Therefore, Liouville's theorem concludes that both sides of \eqref{ParaA-WHE-B-final} are equivalently zero, and then we obtain the solution for $B^+(z)$,
\begin{align}\label{ParaA-B-WHEsol}
B^+(z)&=\frac{\cos\left(\frac{kh}{2}\sin(\theta_{\text{I}})\right)}{K_B^+(z)K_B^-(e^{-iks\cos(\theta_{\text{I}})})(ze^{iks\cos(\theta_{\text{I}})}-1)},
\end{align}
For the second Wiener--Hopf equation in \eqref{ParaA-WHE-BC}, the technique works in exactly the same way and yields the solution,
\begin{align}\label{ParaA-C-WHEsol}
C^+(z)&=-\frac{i\sin\left(\frac{kh}{2}\sin(\theta_{\text{I}})\right)}{K_C^+(z)K_C^-(e^{-iks\cos(\theta_{\text{I}})})(ze^{iks\cos(\theta_{\text{I}})}-1)}.
\end{align}

\subsection{Determining the scattering coefficients}\label{sec:invZT}
Applying the inverse Z--transform to \eqref{ParaA-B-WHEsol} will give,
\begin{align}\label{ParaA-B-sol-int}
B_m=\frac{1}{2\pi i}\frac{\cos\left(\frac{kh}{2}\sin(\theta_{\text{I}})\right)}{K_B^-(e^{-iks\cos(\theta_{\text{I}})})}\oint_{\mathcal{C}} \frac{z^{-m-1}}{K_B^+(z)(ze^{iks\cos(\theta_{\text{I}})}-1)}\text{d}z,
\end{align}
The integration contour $\mathcal{C}$ in \eqref{ParaA-B-sol-int} takes the path that runs anticlockwise around the unit circle (while navigating radially below any singularities on the unit circle) with $z=0$ being the only potential pole inside. To proceed, we must define the expansion of $1/K_B^+(z)$
\begin{align}\label{ParaA-1/KB+-series}
\frac{1}{K_B^+(z)}=\sum_{n=0}^\infty \lambda_{B,n} z^n,
\end{align}
where the coefficients $\lambda_{B,n}$ are defined by 
\begin{align}
\lambda_{B,n}=\frac{1}{n!}\Der{^n}{z^n}\left[\frac{1}{K_B^+(z)}\right]_{z=0}=\frac{1}{2\pi i}\oint_{\mathcal{C}}\frac{z^{-n-1}}{K_B^+(z)}\mathd z.
\end{align}
Then we apply \eqref{ParaA-1/KB+-series} to \eqref{ParaA-B-sol-int}
\begin{align}\label{ParaA-B-sol-int2}
B_m=&\frac{1}{2\pi i}\frac{\cos\left(\frac{kh}{2}\sin(\theta_{\text{I}})\right)}{K_B^-(e^{-iks\cos(\theta_{\text{I}})})} \sum_{n=0}^\infty\oint_{\mathcal{C}} \frac{\lambda_{B,n}z^{n-m-1}}{(ze^{iks\cos(\theta_{\text{I}})}-1)}\text{d}z.
\end{align}
This integral is only non-zero if $m\leq n$. Therefore the evaluation of the $B_m$ coefficients is given by,
\begin{align}\label{ParaA-B-sol}
B_m=&-\frac{\cos\left(\frac{kh}{2}\sin(\theta_{\text{I}})\right)}{K_B^-(e^{-iks\cos(\theta_{\text{I}})})}
\sum_{n=0}^m \left[\lambda_{B,n}e^{iks(m-n)\cos(\theta_{\textrm{I}})}\right].
\end{align}
If we repeat the same procedure for $C_m$, then we obtain,
\begin{align}\label{ParaA-C-sol}
C_m=&\frac{i\sin\left(\frac{kh}{2}\sin(\theta_{\text{I}})\right)}{K_C^-(e^{-iks\cos(\theta_{\text{I}})})}
\sum_{n=0}^m \left[\lambda_{C,n}e^{iks(m-n)\cos(\theta_{\textrm{I}})}\right],
\end{align}
with the coefficients $\lambda_{C,n}$ being derived from the Taylor's series,
\begin{align}\label{ParaA-1/KC+-series}
\frac{1}{K_C^+(z)}=\sum_{n=0}^\infty \lambda_{C,n} z^n,\ \ \text{where}\ \ \lambda_{C,n}=\frac{1}{n!}\Der{^n}{z^n}\left[\frac{1}{K_C^+(z)}\right]_{z=0}=\frac{1}{2\pi i}\oint_{\mathcal{C}}\frac{z^{-n-1}}{K_C^+(z)}\mathd z.
\end{align}
Finally, we find the scattering coefficients using the formulae $A^{(1)}_m=B_m+C_m$ and $A^{(2)}_m=B_m-C_m$ and substitute into \eqref{ParaA-A-gensol} for the scattered field and the total field.

\section{Numerical results}\label{sec:results}
Now that we have the solution for the scattering coefficients of the parallel array problem, we will note some interesting properties and alternative formats. For example, both \eqref{ParaA-B-sol} and \eqref{ParaA-C-sol} can be written as a recurrence relation,
\begin{align}\label{ParaA-B-recur}
B_m=e^{iks\cos(\theta_{\textrm{I}})}B_{m-1}+\frac{\lambda_{B,m}}{\lambda_{B,0}}B_0,\qquad
B_0=&-\frac{\cos\left(\frac{kh}{2}\sin(\theta_{\text{I}})\right)}{K_B^-(e^{-iks\cos(\theta_{\text{I}})})}\lambda_{B,0},\\
\label{ParaA-C-recur}
C_m=e^{iks\cos(\theta_{\textrm{I}})}C_{m-1}+\frac{\lambda_{C,m}}{\lambda_{C,0}}C_0,\qquad
C_0=&\frac{i\sin\left(\frac{kh}{2}\sin(\theta_{\text{I}})\right)}{K_C^-(e^{-iks\cos(\theta_{\text{I}})})}\lambda_{C,0}.
\end{align}
These recurrence relations show that each successive scattering coefficient can be calculated directly from the previous coefficient and the initial coefficient, therefore are a more efficient method of calculation. and help highlight their long term behaviour as $m\rightarrow\infty$. Hence, $B_m$ and $C_m$ have the following leading order behaviour,
\begin{align}\label{ParaA-B-inf}
B_m\sim&-\frac{\cos\left(\frac{kh}{2}\sin(\theta_{\text{I}})\right)}{K_B(e^{-iks\cos(\theta_{\text{I}})})}e^{iksm\cos(\theta_{\textrm{I}})},\\
\label{ParaA-C-inf}
C_m\sim&\frac{i\sin\left(\frac{kh}{2}\sin(\theta_{\text{I}})\right)}{K_C(e^{-iks\cos(\theta_{\text{I}})})}e^{iksm\cos(\theta_{\textrm{I}})},
\end{align}
which allows us to work out the long term behaviour of the scattering coefficients,
\begin{align}
\label{ParaA-A1-inf}A^{(1)}_m\sim&-\left[\frac{e^{-\frac{ikh}{2}\sin(\theta_{\text{I}})}K(e^{-iks\cos(\theta_{\text{I}})})-e^{\frac{ikh}{2}\sin(\theta_{\text{I}})}L(e^{-iks\cos(\theta_{\text{I}})})}{(K(e^{-iks\cos(\theta_{\text{I}})}))^2-(L(e^{-iks\cos(\theta_{\text{I}})}))^2}\right]e^{iksm\cos(\theta_{\textrm{I}})},\\
\label{ParaA-A2-inf}A^{(2)}_m\sim&-\left[\frac{e^{\frac{ikh}{2}\sin(\theta_{\text{I}})}K(e^{-iks\cos(\theta_{\text{I}})})-e^{-\frac{ikh}{2}\sin(\theta_{\text{I}})}L(e^{-iks\cos(\theta_{\text{I}})})}{(K(e^{-iks\cos(\theta_{\text{I}})}))^2-(L(e^{-iks\cos(\theta_{\text{I}})}))^2}\right]e^{iksm\cos(\theta_{\textrm{I}})}.
\end{align}
This long-term behaviour matches perfectly as expected to the solution from the associated problem with two parallel infinite periodic arrays, away from the influence of the termination at the leading edge. Additionally, a sufficiently wide separation compared to the array spacing (i.e. $h\gg s$) will imply that $L(e^{-iks\cos(\theta_{\text{I}})})\ll1$ which leads to each set of the scattering coefficients matching to the solution of the original semi-infinite array problem as seen in \cite{HillsKarp1965} but with an extra factor due to different starting positions. 

Both arrays have the same conditions for resonance (or Wood's anomalies),
$$\frac{ks}{2\pi}(1\mp\cos(\theta_{\textrm{I}}))\in\ZZ.$$
These conditions include resonance caused by the incident wave as well as any waves transmitted through one array onto the other and being reflected back and forth between the two arrays. Similarly to single semi-infinite and infinite array problems, increasing the quantity $ks$ will take it past resonance and double resonance points which add extra reflected and transmitted waves to the scattered wave field.

Another interesting quantity is $kh\sin(\theta_{\textrm{I}})$ which determines the relative phase of the incident wave at the two arrays, and consequently how symmetric or antisymmetric the resulting scattered field will be. Special cases include, if $kh\sin(\theta_{\textrm{I}})$ is an even multiple of $\pi$ then $C_m=0$ for all $m$ and the scattered field is perfectly symmetric and if $kh\sin(\theta_{\textrm{I}})$ is an odd multiple of $\pi$, then $B_m=0$ for all $m$ and the scattered field is antisymmetric. In the first case, the incident wave excites the two arrays in phase, whereas in the second it excites them out of phase and consequently, one of the two decoupled Wiener–Hopf problems disappears entirely.
\begin{figure}[ht]\centering
\includegraphics[width=0.685\textwidth]{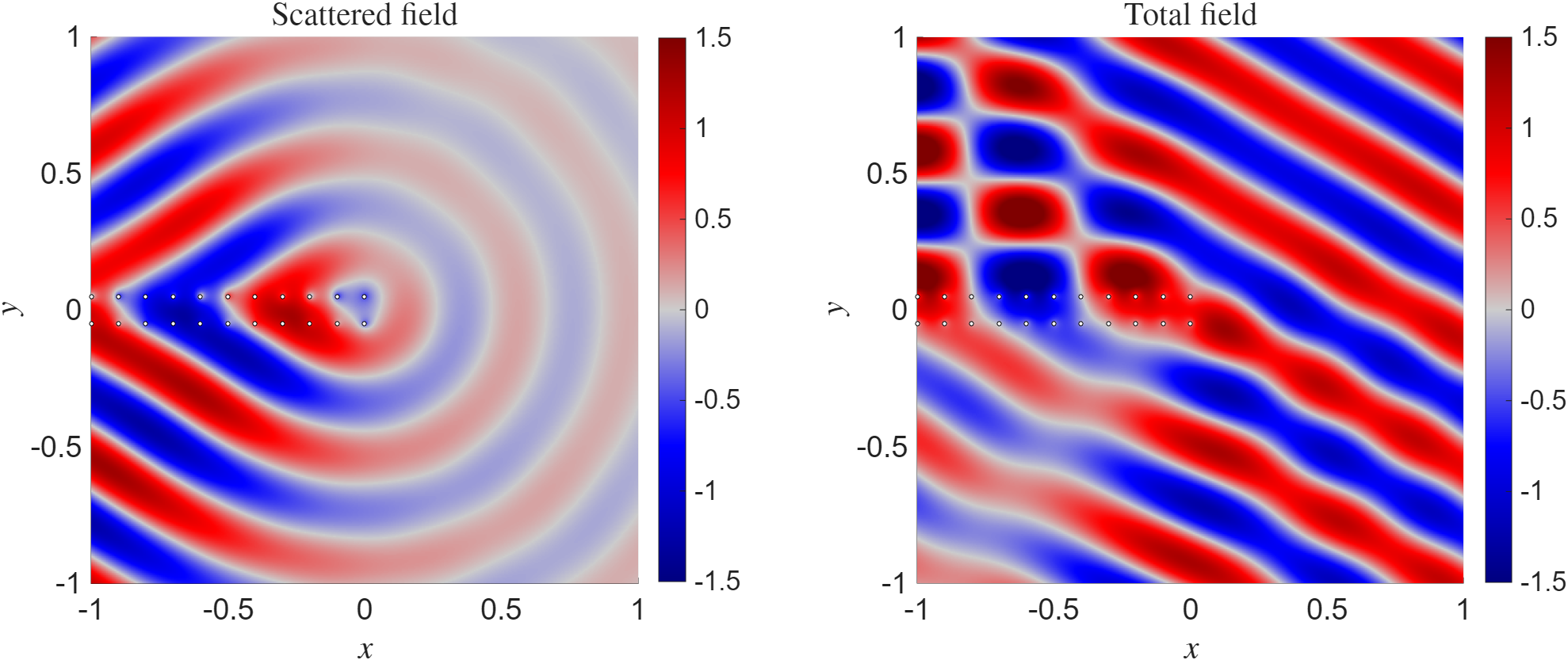}\\
\includegraphics[width=0.685\textwidth]{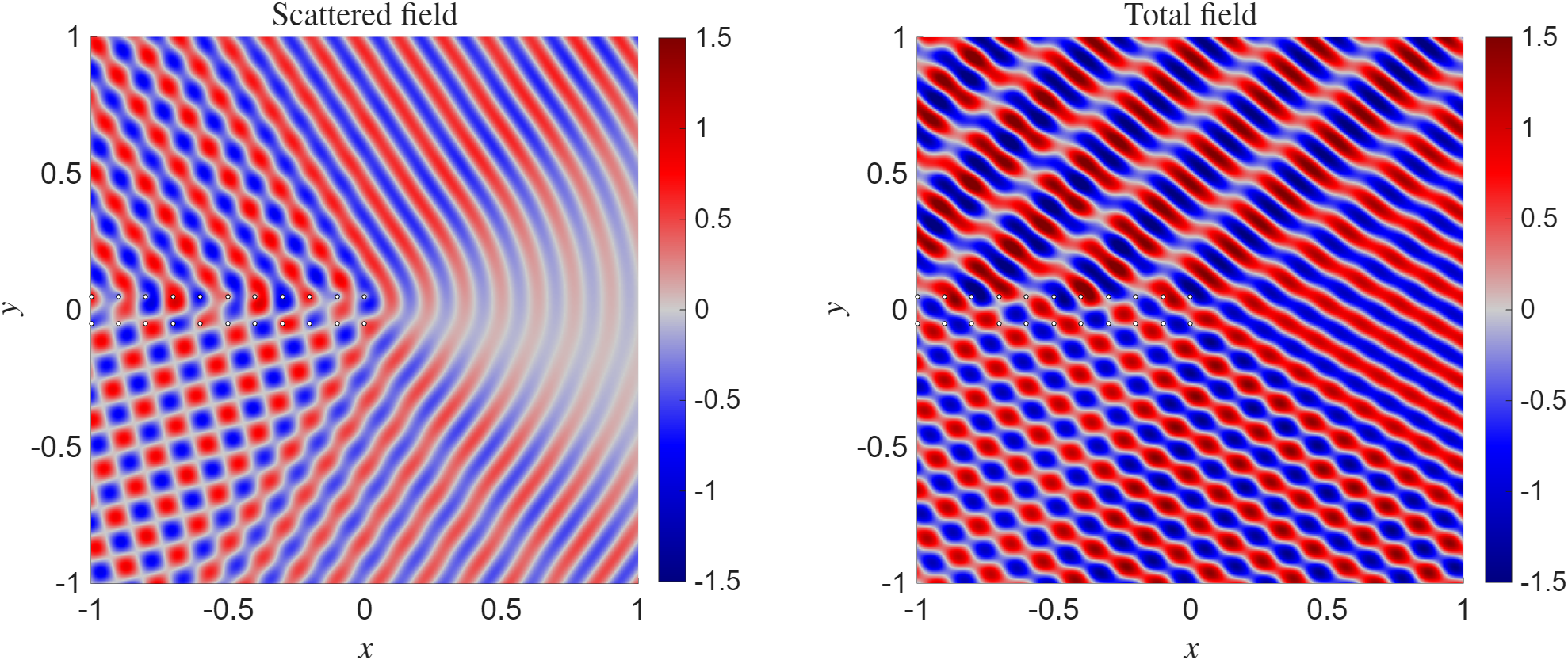}\\
\includegraphics[width=0.685\textwidth]{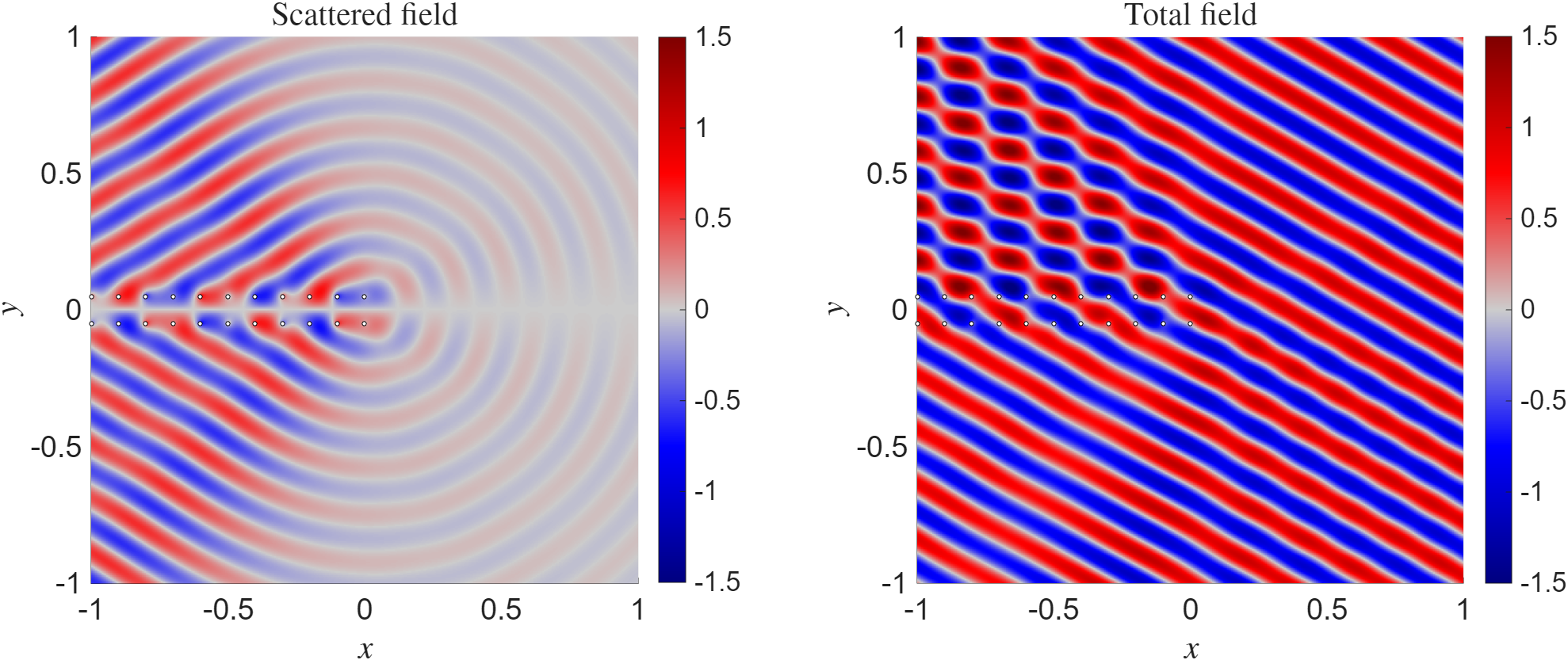}
\caption{Real part of the scattered field $\Phi_{\textrm{S}}$ (left) and the total field $\Phi$ (right). Here, the parameters are $a=0.001$, $s=h=0.1$ and $\theta_{\textrm{I}}=\pi/3$ with different wavenumbers $k=5\pi$, $15\pi$ and $k=20\pi/\sqrt{3}$ for the top, middle and bottom row respectively.}
\label{fig:Phi_test_case}
\end{figure}

Figure \ref{fig:Phi_test_case} plots the real part of the scattered field $\Phi_{\textrm{S}}$ (left) and the total field $\Phi$ (right) for two test cases using \eqref{ParaA-A-gensol} truncated at 1000 scatterers per array. The first test case has parameters; $a=0.001$, $s=h=0.1$, $k=5\pi$ and $\theta_{\textrm{I}}=\pi/3$. The second test case is similar but with a higher wavenumber $k=15\pi$ instead. The third case has a specific wavenumber $k=20\pi/\sqrt{3}$ which leads to the symmetric coefficients $B_m$ being identically zero for all $m$ giving us a fully antisymmetric scattered field. Between all these cases, the first and third ones have both resonance conditions within 
$$0<\frac{ks}{2\pi}(1\mp\cos(\theta_{\textrm{I}}))<1$$
which implies that there is only one reflected and transmitted wave per array. For the second case however, one of the conditions is past the resonance point
$$0<\frac{ks}{2\pi}(1-\cos(\theta_{\textrm{I}}))<1<\frac{ks}{2\pi}(1+\cos(\theta_{\textrm{I}})),$$
which means that it has two reflected and transmitted waves per array.

\begin{figure}[ht]\centering
\includegraphics[width=0.45\textwidth]{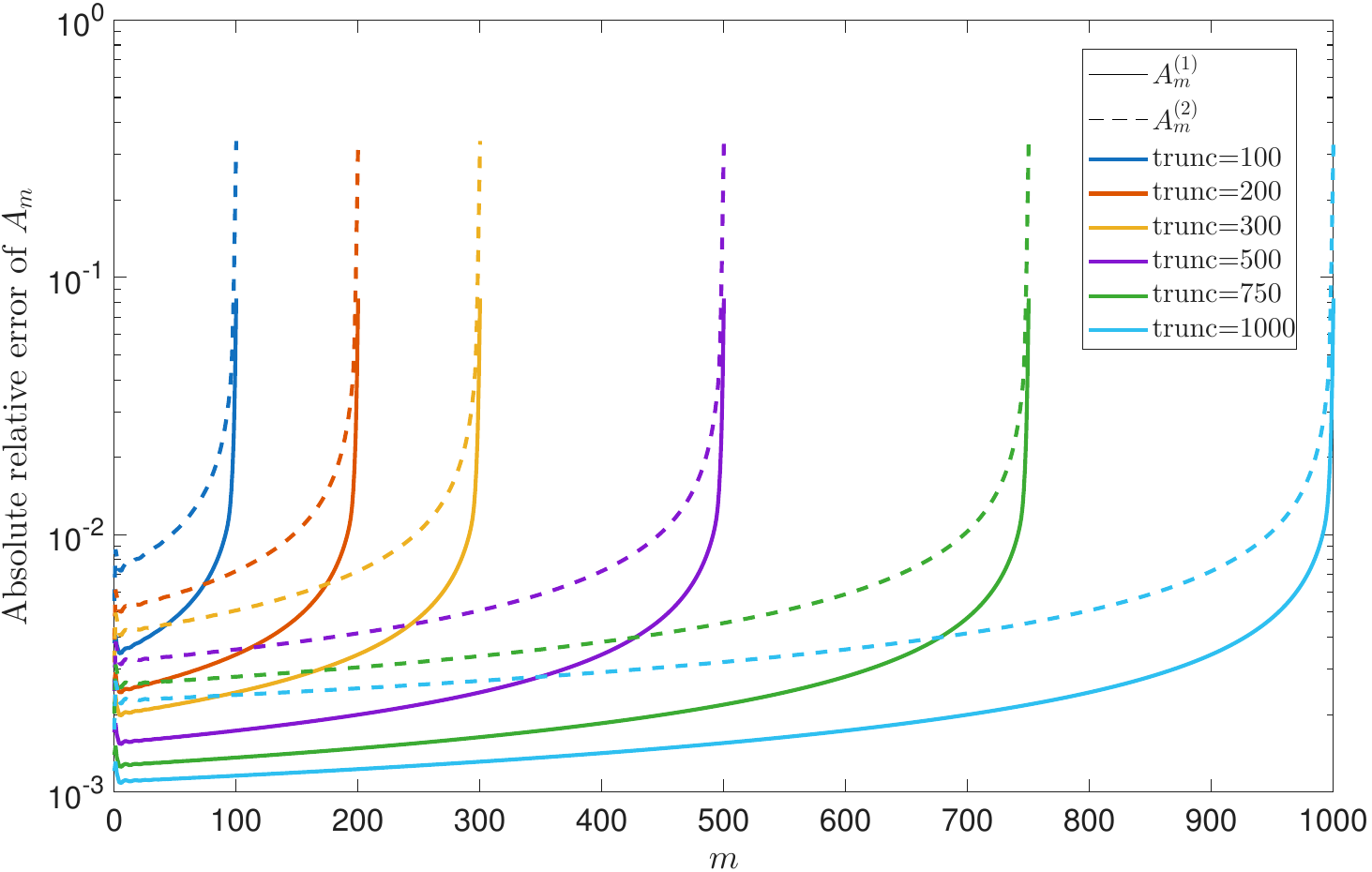}
\includegraphics[width=0.45\textwidth]{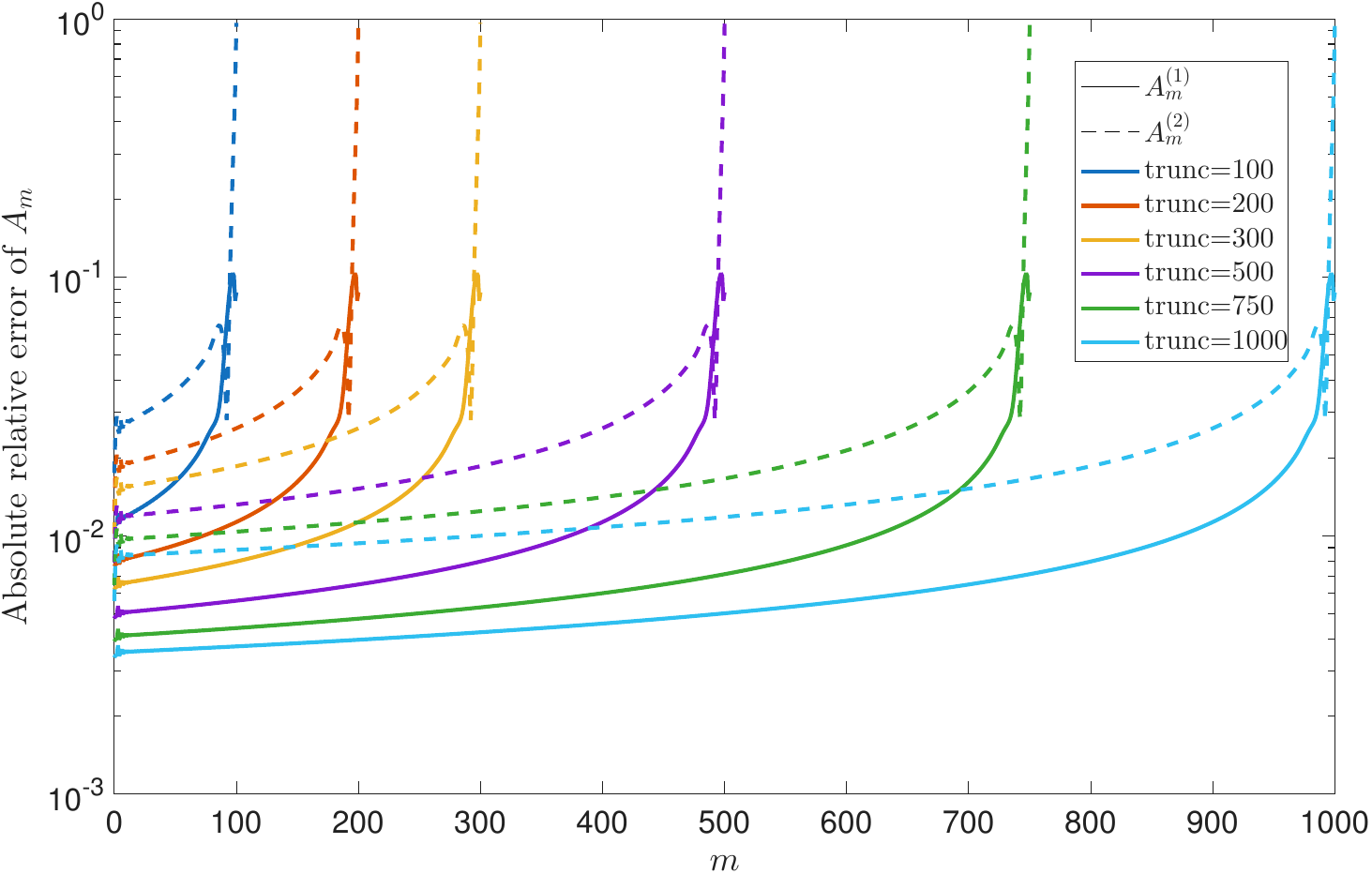}\\
\includegraphics[width=0.45\textwidth]{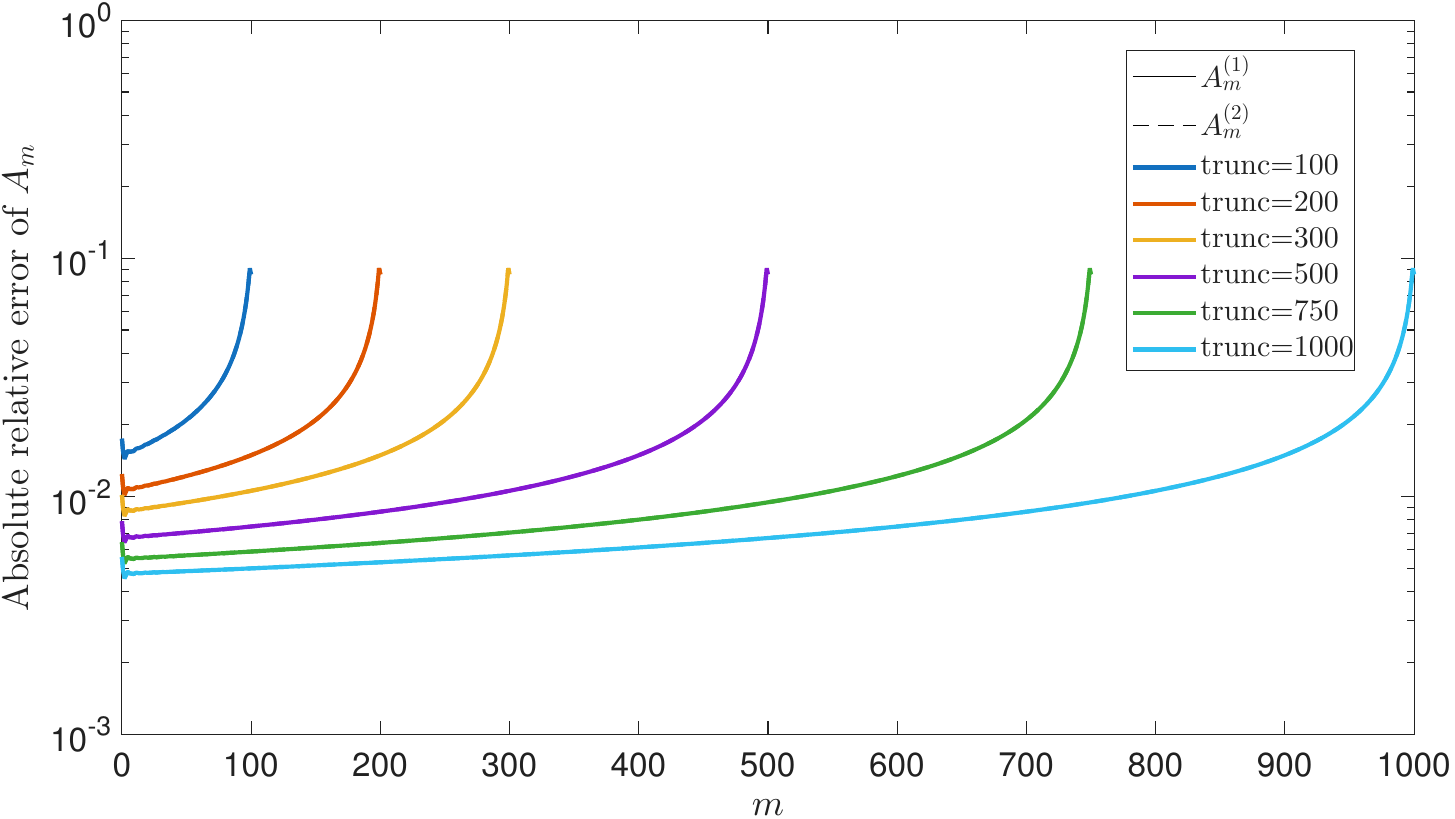}
\caption{Absolute relative error of the scattering coefficients calculated in \citep{MSIApaper} versus the exact solution present in this article. Solid lines look at the coefficients of the top array $A^{(1)}_m$, dashed lines look at the bottom array $A^{(2)}_m$ and different colours represent different truncations used in the method of \citep{MSIApaper}. Here, we reuse the test cases from Figure \ref{fig:Phi_test_case} with parameters $a=0.001$, $s=h=0.1$ and $\theta_{\textrm{I}}=\pi/3$ with different wavenumbers $k=5\pi$ and $15\pi$ for the left and right side on the top row and $k=20\pi/\sqrt{3}$ for the bottom row respectively.}
\label{fig:MSIA_error}
\end{figure}

A major advantage to the results of this article is that, to the author's knowledge, this is the only multiple semi-infinite array scenario for which an exact solution has been found. This makes it the ideal problem to test the validity and accuracy of the method outlined in \citep{MSIApaper} for which before now we have only compared with other numerical or approximate methods. Figure \ref{fig:MSIA_error} demonstrates this by displaying the absolute relative error of the scattering coefficients for the earlier test cases from Figure \ref{fig:Phi_test_case}. In these plots, solid lines look at the coefficients of the top array $A^{(1)}_m$, dashed lines look at the bottom array $A^{(2)}_m$. The different colours represent different truncations used in the method of \citep{MSIApaper}. Here, the parameters are given by $a=0.001$, $s=h=0.1$ and $\theta_{\textrm{I}}=\pi/3$ with different wavenumbers $k=5\pi$ and $15\pi$ for the left and right side on the top row and $k=20\pi/\sqrt{3}$ for the bottom row respectively. The last plot is considering the antisymmetric example so the solid and dashed lines are on top of each other because $A^{(1)}_m$ and $A^{(1)}_m$ differ only by a minus sign. The key conclusion of this figure is to show that for this parallel semi-infinite array problem, the method of \citep{MSIApaper} is clearly converging for all test cases as the truncation increases with the largest errors located on the truncated end. It also shows that \citep{MSIApaper} calculates the coefficients of the first array encountered by the incident wave better than the other array.

\section{Conclusions}
In this article, we derived an exact Wiener-Hopf solution for acoustic scattering by two identical, parallel semi-infinite arrays of point scatterers. The symmetry of the configuration allows the coupled matrix Wiener--Hopf system to be decomposed into independent even and odd problems, with factorisable kernels given by $K(z)+L(z)$ and $K(z)-L(z)$, respectively. The standard Wiener--Hopf procedure then provides explicit expressions and recurrence relations for the corresponding symmetric and antisymmetric components of the scattering coefficients, from which the scattering response of both arrays can be recovered. The analysis demonstrates that the exchange symmetry of this two-array configuration provides a simple and distinctive way of avoiding the difficulties of general matrix Wiener--Hopf factorisation. Additionally, the kernel functions admit several different evaluation methods, providing flexibility in their practical computation. The resulting formulation provides an exact analytical treatment of this coupled semi-infinite array problem and offers a useful special case within the broader class of multiple semi-infinite array scattering problems. In particular, the exact solution provides a benchmark against which more general numerical and approximate approaches to multiple array scattering, such as the method developed in \citep{MSIApaper}, can be assessed. Figure \ref{fig:MSIA_error} demonstrates that the methodology of \citep{MSIApaper} converges for the test cases considered as the truncation increases, with the largest errors occurring near the truncated ends of the arrays.

A natural extension would be to consider two parallel arrays whose starting positions are offset horizontally, forming a staggered configuration. This would be the array equivalent of the staggered plate problem presented in \cite{Abrahams1988,Abrahams1988part2}. In this case, the relative displacement modifies the inter-array interactions and hence the off-diagonal entries of the matrix Wiener–Hopf kernel. The exchange symmetry exploited in the present work is then lost because the symmetry relation $L(z)=L(1/z)$ would not apply anymore. Therefore, the decomposition into independent symmetric and antisymmetric problems is no longer applicable. For offsets that are integer multiples of the array spacing, the modified interactions may instead be related to the original lattice sums through shifts in the lattice index, introducing additional powers of the transform variable into the matrix kernel. For example, 
\begin{align*}
\begin{pmatrix}
K(z)&z^{-N}L(z)\\z^NL(z)&K(z)
\end{pmatrix}
\end{align*}
where the offset between the two arrays is $Ns$ in the positive $x$ direction. Understanding whether this structure can be exploited to obtain an explicit factorisation would provide a natural direction for extending the present analysis. More generally, such configurations would help establish the extent to which the geometrical structure can be used to simplify matrix Wiener–Hopf problems for multiple semi-infinite arrays.


\appendix
\section{Practical evaluation of Wiener--Hopf kernels}\label{sec:Kernels}
\begin{figure}[ht]\centering
\includegraphics[width=0.7\textwidth]{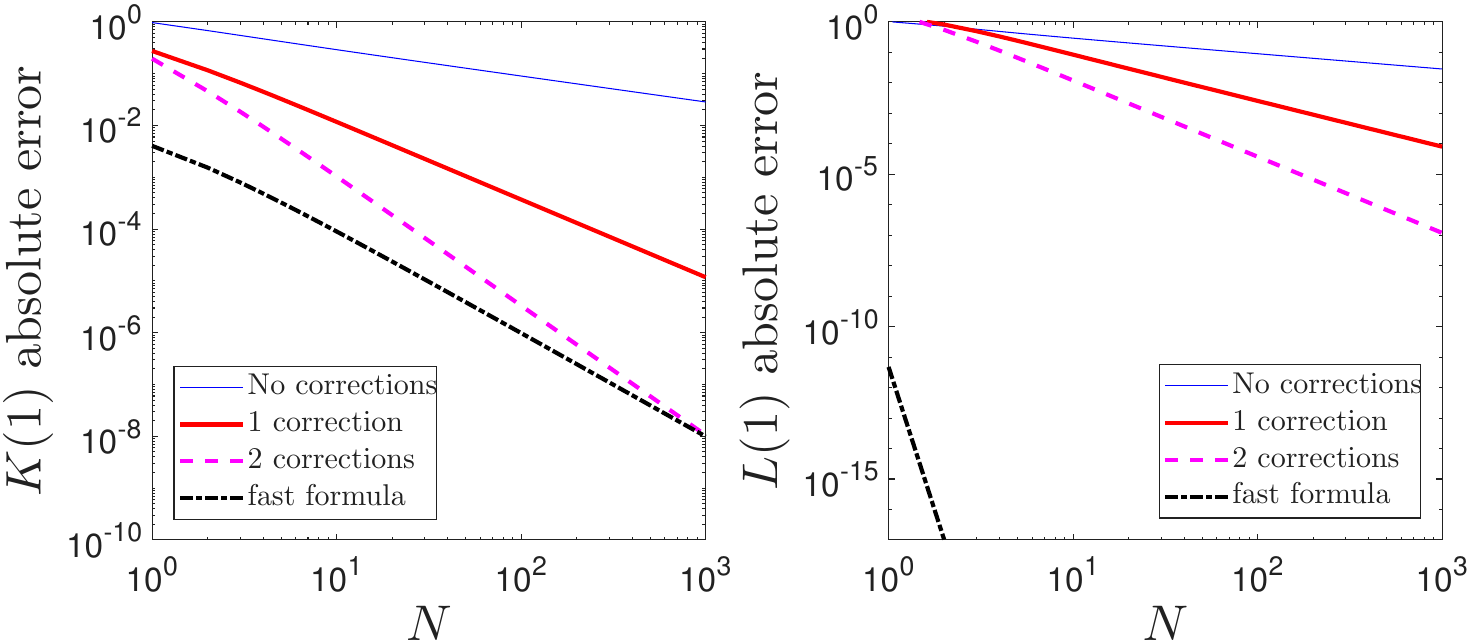}
\caption{Absolute error of $K(z)$ (left) and $L(z)$ (right) against truncation $N$ with the blue, red, magenta and black lines using the original formulae \eqref{ParaA-K(z)}-\eqref{ParaA-L(z)} with 0, 1 and 2 tail-end corrections and the alternative formulae \eqref{Kernels-K(z)-fast}-\eqref{Kernels-L(z)-fast} respectively. Here, the parameters are $a=0.001$, $s=h=0.1$ and $k=5\pi$. We find that as $N\rightarrow\infty$, the error converges like $O(N^{-1/2})$, $O(N^{-3/2})$ and $O(N^{-5/2})$ for the blue, red and magenta lines respectively. Additionally, we find that as $N\rightarrow\infty$, the black line of the left plot indicates $O(N^{-2})$ convergence and the right one is exponential.}
\label{fig:KL_comparison}
\end{figure}

We discussed the practical evaluation of $K(z)$ in the appendix of \citep{HWpaperI} because it is a slow-convergent series. This included the use of the method of tail-end asymptotics as inspired by \citep{Lynott_19}. While this method is quite effective at accelerating the convergence of \eqref{ParaA-K(z)}, the sum is still only able to converge on the unit circle $|z|=1$. Alternatively, we can rewrite $K(z)$ into a fast convergent series which requires the mapping $z=e^{it}$, 
\begin{align}\label{Kernels-K(z)-fast}
K(e^{it})=&\ H^{(1)}_0(ka)-1-\frac{2i}{\pi}\left(\gamma+\ln\left(\frac{ks}{4\pi}\right)\right)+\frac{2}{\kappa(t)}
+\sum_{\substack{\ell=-\infty\\\ell\neq 0}}^\infty\left[\frac{2}{\kappa(t-2\pi\ell)}+\frac{i}{\pi |\ell|}\right],
\end{align}
where $\kappa(\tau)=\sqrt{(ks)^2-\tau^2}$ and $\gamma=0.5772...$ is the Euler–Mascheroni constant. In this formula, there are two sets of branch cuts starting at the points $\tau=\pm ks$. The ones starting at $\tau=ks$ (resp. $\tau=-ks$) are orientated vertically towards positive (resp. negative) imaginary infinity in the $\tau$ complex plane. All of the square roots in \eqref{Kernels-K(z)-fast} are defined to ensure that $\IM{\kappa(\tau)}\geq0$ when $\tau$ is real. In other words, the branches are selected such that $\kappa(0)=ks$. This formula offers two advantages over the original: it converges more rapidly, improving the convergence rate from $O(N^{-1/2})$ to $O(N^{-2})$, where $N$ denotes the truncation parameter, and it admits further acceleration, with each correction improving the convergence rate by an additional factor of $N^{-2}$ (as shown in \citep{HWpaperI}). Furthermore, the formula can converge beyond the unit circle. The left-hand side of Figure \ref{fig:KL_comparison} compares $K(z)$ obtained using several formulae, including \eqref{ParaA-K(z)} with zero, one, and two tail-end corrections as well as \eqref{Kernels-K(z)-fast}, by plotting the absolute error for the test point $z=1$.

The additional kernel $L(z)$ has many similar properties to $K(z)$ including the original definition being a slow-convergent series with the method of tail-end asymptotics being applicable to accelerate convergence (improving the convergence rate from $O(N^{-1/2})$ to $O(N^{-2})$). It also has it's own fast convergent formula which can be found by substituting Hankel's integral formula into \eqref{ParaA-L(z)} given by,
$$H^{(1)}_0(\sqrt{x^2+y^2})=\frac{1}{\pi}\int_{-\infty}^{\infty}\frac{e^{ix\zeta+i|y|\sqrt{1-\zeta^2}}}{\sqrt{1-\zeta^2}}\mathd \zeta,$$
where we put $x=ksn$ and $y=kh$, and use Poisson's summation formula to obtain,
\begin{align}\label{Kernels-L(z)-fast}
L(e^{it})=2\sum_{\ell=-\infty}^{\infty}\frac{e^{\frac{ih}{s}\kappa(t-2\pi\ell)}}{\kappa(t-2\pi\ell)}.
\end{align}
This new sum also extends the region of convergence beyond the unit circle and is exponentially convergent. The right side of Figure \ref{fig:KL_comparison} is a $L(z)$ comparison between different formulae including \eqref{ParaA-L(z)} with 0, 1 and 2 tail-end corrections as well as \eqref{Kernels-L(z)-fast} by plotting the absolute error for a test point $z=1$. In this article, we exclusively use the formulae \eqref{Kernels-K(z)-fast} and \eqref{Kernels-L(z)-fast} for analytical evaluations of the kernels $K(z)$ and $L(z)$ respectively. We also reuse the ``AAA" approximation by Chebfun in MATLAB to create rational approximations of $K_B(z)$ and $K_C(z)$ (see \citep{HWpaperI} for details) for efficient evaluations of the coefficients $\lambda_{B,n}$ and $\lambda_{C,n}$ and then the evaluation of the scattering coefficients $A^{(1)}_m$ and $A^{(2)}_m$ is relatively fast to compute.


\bibliographystyle{abbrvnat}
\bibliography{HWbib}

@article{HWpaperI,
author = {Nethercote, Matthew Allan and Kisil, Anastasia V. and Assier, Rapha{\"{e}}l C.},
title = {{Diffraction of acoustic waves by a wedge of point scatterers}},
journal = {SIAM J. Appl. Math.},
fjournal = {SIAM Journal on Applied Mathematics},
volume = {82},
number = {3},
pages = {872-898},
year = {2022}
}

@article{ArrayResonanceFoldyPaper,
author = {Nethercote, Matthew Allan and Thompson, Ian and Kisil, Anastasia V. and Assier, Rapha{\"{e}}l C.},
journal = {Proc. R. Soc. A},
fjournal = {Proceedings of the Royal Society of London. A. Mathematical and Physical Sciences},
number = {20220604},
title = {{Array scattering resonance in the context of Foldy's approximation}},
volume = {478},
year = {2022}
}

@article{MSIApaper,
author = {Nethercote, Matthew Allan and Kisil, Anastasia V. and Assier, Rapha{\"{e}}l C.},
doi = {10.1121/10.0020844},
journal = {J. Acoust. Soc. Am.},
fjournal = {Journal of Acoustic Society of America},
title = {{Diffraction of acoustic waves by multiple semi-infinite arrays}},
url = {https://doi.org/10.1121/10.0020844},
volume = {154},
number = {3},
pages = {1493--1504},
year = {2023}
}

@article{Chremmos2011,
author = {Chremmos, Ioannis and Fikioris, George},
journal = {J. Opt. Soc. Am. A},
fjournal = {Journal of the Optical Society of America A},
number = {6},
pages = {962--969},
title = {{Traveling waves in two parallel infinite linear point-scatterer arrays}},
volume = {28},
year = {2011}
}

@article{Abrahams1987a,
author = {Abrahams, Ian David},
journal = {Wave Motion},
number = {4},
pages = {289--300},
title = {{Scattering of sound by two parallel semi-infinite screens}},
volume = {9},
year = {1987}
}

@article{Abrahams1988,
author = {Abrahams, Ian David and Wickham, G. R.},
journal = {Proc. R. Soc. A},
fjournal = {Proceedings of the Royal Society of London. A. Mathematical and Physical Sciences},
number = {1858},
pages = {131--156},
title = {{On the scattering of sound by two semi-infinite parallel staggered plates - I. Explicit matrix Wiener-Hopf factorization}},
volume = {420},
year = {1988}
}

@article{Abrahams1988part2,
author = {Abrahams, Ian David and Wickham, G. R.},
journal = {Proc. R. Soc. A},
fjournal = {Proceedings of the Royal Society of London. A. Mathematical and Physical Sciences},
number = {1872},
pages = {139--171},
title = {{The scattering of sound by two semi-infinite parallel staggered plates. II. Evaluation of the velocity potential for an incident plane wave and an incident duct mode}},
volume = {427},
year = {1990}
}

@article{KisilWHReview2021,
author = {Kisil, Anastasia V. and Abrahams, Ian David and Mishuris, Gennady and Rogosin, Sergei V.},
journal = {Proc. R. Soc. A},
fjournal = {Proceedings of the Royal Society A: Mathematical, Physical and Engineering Sciences},
number = {2254},
title = {{The Wiener-Hopf technique, its generalizations and applications: Constructive and approximate methods}},
volume = {477},
year = {2021}
}

@article{RogosinMishuris2016,
author = {Rogosin, Sergei V. and Mishuris, Gennady S.},
journal = {IMA J. Appl. Math.},
fjournal = {IMA Journal of Applied Mathematics},
number = {2},
pages = {365--391},
title = {{Constructive methods for factorization of matrix-functions}},
volume = {81},
year = {2016}
}

@article{HaslingerJonesMovchan2018,
author = {Haslinger, Stewart G. and Jones, Ian S. and Movchan, Natalia V. and Movchan, Alexander B.},
title = {{Localization in semi-infinite herringbone waveguides}},
journal = {Proc. R. Soc. A},
fjournal = {Proceedings of the Royal Society A: Mathematical, Physical and Engineering Sciences},
volume = {474},
pages = {20170590},
year = {2018}
}

@article{JonesMovchan2017,
author = {Jones, Ian S. and Movchan, Natalia V. and Movchan, Alexander B.},
title = {{Blockage and guiding of flexural waves in a semi-infinite double grating}},
journal = {Mathematical Methods in the Applied Sciences},
volume = {40},
number = {9},
pages = {3265--3282},
year = {2017}
}

@article{Heins1948part1,
author = {Heins, Albert E.},
journal = {Q. Appl. Math.},
fjournal = {Quarterly of Applied Mathematics},
number = {2},
pages = {157--166},
title = {{The radiation and transmission properties of a pair of semi-infinite parallel plates - I}},
volume = {6},
year = {1948}
}

@article{Heins1948part2,
author = {Heins, Albert E.},
journal = {Q. Appl. Math.},
fjournal = {Quarterly of Applied Mathematics},
number = {3},
pages = {215--220},
title = {{The radiation and transmission properties of a pair of parallel plates - II}},
volume = {6},
year = {1948}
}

@article{Linton1998,
author = {Linton, Chris M.},
journal = {J. Eng. Math.},
fjournal = {Journal of Engineering Mathematics},
number = {4},
pages = {377--402},
title = {{The {G}reen's function for the two-dimensional {H}elmholtz equation in periodic domains}},
volume = {33},
year = {1998},
}

@article{Linton2006,
author = {Linton, Chris M.},
journal = {J Phys. A},
fjournal = {Journal of Physics A: Mathematical and General},
number = {13},
pages = {3325--3339},
title = {{Schl{\"{o}}milch series that arise in diffraction theory and their efficient computation}},
volume = {39},
year = {2006}
}

@article{Linton2010,
author = {Linton, Chris M.},
journal = {SIAM Rev.},
fjournal = {SIAM Review},
number = {4},
pages = {630--674},
title = {{Lattice sums for the {H}elmhoitz equation}},
volume = {52},
year = {2010},
}

@article{KrynkinMcIver2009,
author = {Krynkin, A. and McIver, P.},
journal = {Waves Random Complex Media},
fjournal = {Waves in Random and Complex Media},
number = {2},
pages = {347--365},
title = {{Approximations to wave propagation through a lattice of Dirichlet scatterers}},
volume = {19},
year = {2009}
}

@article{WienerHopf1931,
author = {Wiener, Norbert and Hopf, Eberhard},
journal = {Sitzungsberichte der Preussischen Akademie der Wissenschaften, Physikalisch-Mathematische Klasse (in German )},
pages = {696--706},
title = {{{\"{U}}ber eine klasse singul{\"{a}}rer integralgleichungen}},
volume = {31},
year = {1931}
}

@article{HillsKarp1965,
author = {Hills, Norman L. and Karp, Samuel N.},
journal = {Comm. Pure Appl. Math},
fjournal = {Communications on Pure and Applied Mathematics},
pages = {203--233},
title = {{Semi-Infinite Diffraction Gratings-I}},
volume = {XVIII},
year = {1965}
}

@article{LintonMartin2004,
author = {Linton, Chris M. and Martin, Paul A.},
journal = {SIAM J. Appl. Math.},
fjournal = {SIAM Journal on Applied Mathematics},
number = {3},
pages = {1035--1056},
title = {{Semi-Infinite Arrays of Isotropic Point Scatterers. A Unified Approach}},
volume = {64},
year = {2004}
}

@article{LintonPorterThompson2007,
author = {Linton, Chris M. and Porter, R. and Thompson, Ian},
journal = {SIAM J. Appl. Math.},
fjournal = {SIAM Journal on Applied Mathematics},
number = {5},
pages = {1233--1258},
title = {{Scattering by a semi-infinite periodic array and the excitation of surface waves}},
volume = {67},
year = {2007}
}

@article{Movchan2009,
author = {Movchan, Natalia V. and McPhedran, R. C. and Movchan, Alexander B. and Poulton, C. G.},
journal = {Proc. R. Soc. A},
fjournal = {Proceedings of the Royal Society A: Mathematical, Physical and Engineering Sciences},
number = {2111},
pages = {3383--3400},
title = {{Wave scattering by platonic grating stacks}},
volume = {465},
year = {2009}
}

@article{Haslinger2014,
author = {Haslinger, Stewart G. and Movchan, Alexander B. and Movchan, Natalia V. and McPhedran, R. C.},
journal = {Waves Random Complex Media},
fjournal = {Waves in Random and Complex Media},
number = {2},
pages = {126--148},
title = {{Symmetry and resonant modes in platonic grating stacks}},
volume = {24},
year = {2014}
}

@article{Haslinger2016,
author = {Haslinger, Stewart G. and Craster, Richard V. and Movchan, Alexander B. and Movchan, Natalia V. and Jones, Ian S.},
journal = {Proc. R. Soc. A},
fjournal = {Proceedings of the Royal Society A: Mathematical, Physical and Engineering Sciences},
number = {20152186},
title = {{Dynamic interfacial trapping of flexural waves in structured plates}},
volume = {472},
year = {2016}
}

@article{Foldy1945,
author = {Foldy, Leslie L.},
journal = {Phys. Rev.},
fjournal = {Physical Review},
number = {3-4},
pages = {107--119},
title = {{The multiple scattering of waves. I. General theory of isotropic scattering by randomly distributed scatterers}},
volume = {67},
year = {1945}
}

@article {Lynott_19,
AUTHOR = {Lynott, Georgia M. and Andrew, Victoria and Abrahams, I. David and Simon, Michael J. and Parnell, William J. and Assier, Rapha\"{e}l C.},
TITLE = {Acoustic scattering from a one-dimensional array; tail-end asymptotics for efficient evaluation of the quasi-periodic Green's function},
JOURNAL = {Wave Motion},
FJOURNAL = {Wave Motion. An International Journal Reporting Research on Wave Phenomena},
VOLUME = {89},
YEAR = {2019},
PAGES = {232--244},
}

@article{LawrieAbrahams2007,
author = {Lawrie, Jane B. and Abrahams, Ian David},
journal = {J. Eng. Math.},
fjournal = {Journal of Engineering Mathematics},
number = {4},
pages = {351--358},
title = {{A brief historical perspective of the Wiener-Hopf technique}},
volume = {59},
year = {2007}
}

\end{document}